%% file: main.tex
\documentclass[sigconf]{acmart}
\usepackage{multirow}
\usepackage{enumitem}
\usepackage{arydshln}
\usepackage{subfigure}		%子图
\usepackage{booktabs}
\usepackage{bbding}
\usepackage{pifont}
\usepackage{fontawesome}

\AtBeginDocument{%
  }
\usepackage{listings}
\usepackage{xcolor}

\copyrightyear{2025} 
\acmYear{2025} 
\setcopyright{acmlicensed}
\acmConference[WWW Companion '25]{Companion
 Proceedings of the ACM Web Conference 2025}{April 28-May 2, 2025}{Sydney,
 NSW, Australia}
 \acmBooktitle{Companion Proceedings of the ACM Web Conference 2025 (WWW
 Companion '25), April 28-May 2, 2025, Sydney, NSW, Australia}
 \acmDOI{10.1145/3701716.3715213}
 \acmISBN{979-8-4007-1331-6/25/04}

\begin{document}

\title{Comment Staytime Prediction with LLM-enhanced Comment Understanding}

\author{Changshuo Zhang}
\orcid{0009-0001-8481-9421}
\authornote{Both authors contributed equally to this research.}
\email{zhangchangshuo@kuaishou.com}
\affiliation{%
  \institution{Kuaishou Technology Co., Ltd.}
  \city{Beijing}
  \country{China}
}

\author{Zihan Lin}
\orcid{0000-0002-6877-4470}
\authornotemark[1]
\email{linzihan03@kuaishou.com}
\affiliation{%
  \institution{Kuaishou Technology Co., Ltd.}
  \city{Beijing}
  \country{China}
}

\author{Shukai Liu}
\orcid{0009-0009-9492-9868}
\authornote{Shukai Liu is the corresponding author.}
\email{liushukai03@kuaishou.com}
\affiliation{%
  \institution{Kuaishou Technology Co., Ltd.}
  \city{Beijing}
  \country{China}
}

\author{Yongqi Liu}
\orcid{0009-0005-2153-6722}
\email{liuyongqi@kuaishou.com}
\affiliation{%
  \institution{Kuaishou Technology Co., Ltd.}
  \city{Beijing}
  \country{China}
}

\author{Han Li}
\orcid{0009-0000-9801-9292}
\email{lihan08@kuaishou.com}
\affiliation{%
  \institution{Kuaishou Technology Co., Ltd.}
  \city{Beijing}
  \country{China}
}

\newcommand{\ourname}{$\text{LCU}$}

\renewcommand{\shortauthors}{Changshuo Zhang, Zihan Lin, Shukai Liu, Yongqi Liu, and Han Li}

\input{sections/0_abs}

\begin{CCSXML}
<ccs2012>
   <concept>
       <concept_id>10002951.10003317.10003347.10003350</concept_id>
       <concept_desc>Information systems~Recommender systems</concept_desc>
       <concept_significance>500</concept_significance>
       </concept>
 </ccs2012>
\end{CCSXML}

\ccsdesc[500]{Information systems~Recommender systems}

\keywords{Staytime Prediction, Large Language Model, Comment Ranking}

\maketitle

\input{sections/1_intro}

\input{sections/3.0_exp}
\input{sections/3_method}
\input{sections/4_exp}
\input{sections/2_related}
\input{sections/5_conc}
\newpage
\bibliographystyle{ACM-Reference-Format}
\bibliography{sample-base}

\appendix
\input{sections/6_app}
\end{document}

%% file: sections/0_abs.tex
\begin{abstract}
In modern online streaming platforms, the comments section plays a critical role in enhancing the overall user experience. Understanding user behavior within the comments section is essential for comprehensive user interest modeling. A key factor of user engagement is \textbf{staytime}, which refers to the amount of time that users browse and post comments. Existing watchtime prediction methods struggle to adapt to staytime prediction, overlooking interactions with individual comments and their interrelation. In this paper, we present a micro-video recommendation dataset with video comments~(named as \textbf{KuaiComt}) which is collected from Kuaishou platform. correspondingly, we propose a practical framework for comment staytime prediction with \textbf{L}LM-enhanced \textbf{C}omment \textbf{U}nderstanding (LCU). Our framework leverages the strong text comprehension capabilities of large language models (LLMs) to understand textual information of comments, while also incorporating fine-grained comment ranking signals as auxiliary tasks. The framework is two-staged: first, the LLM is fine-tuned using domain-specific tasks to bridge the video and the comments; second, we incorporate the LLM outputs into the prediction model and design two comment ranking auxiliary tasks to better understand user preference. Extensive offline experiments demonstrate the effectiveness of our framework, showing significant improvements on the task of comment staytime prediction. Additionally, online A/B testing further validates the practical benefits on industrial scenario. Our dataset KuaiComt~\footnote{https://github.com/lyingCS/KuaiComt.github.io} and code for LCU~\footnote{https://github.com/lyingCS/LCU} are fully released.
\end{abstract}

%% file: sections/1_intro.tex
\begin{figure}[t]
\centering
\includegraphics[width=0.47 \textwidth]{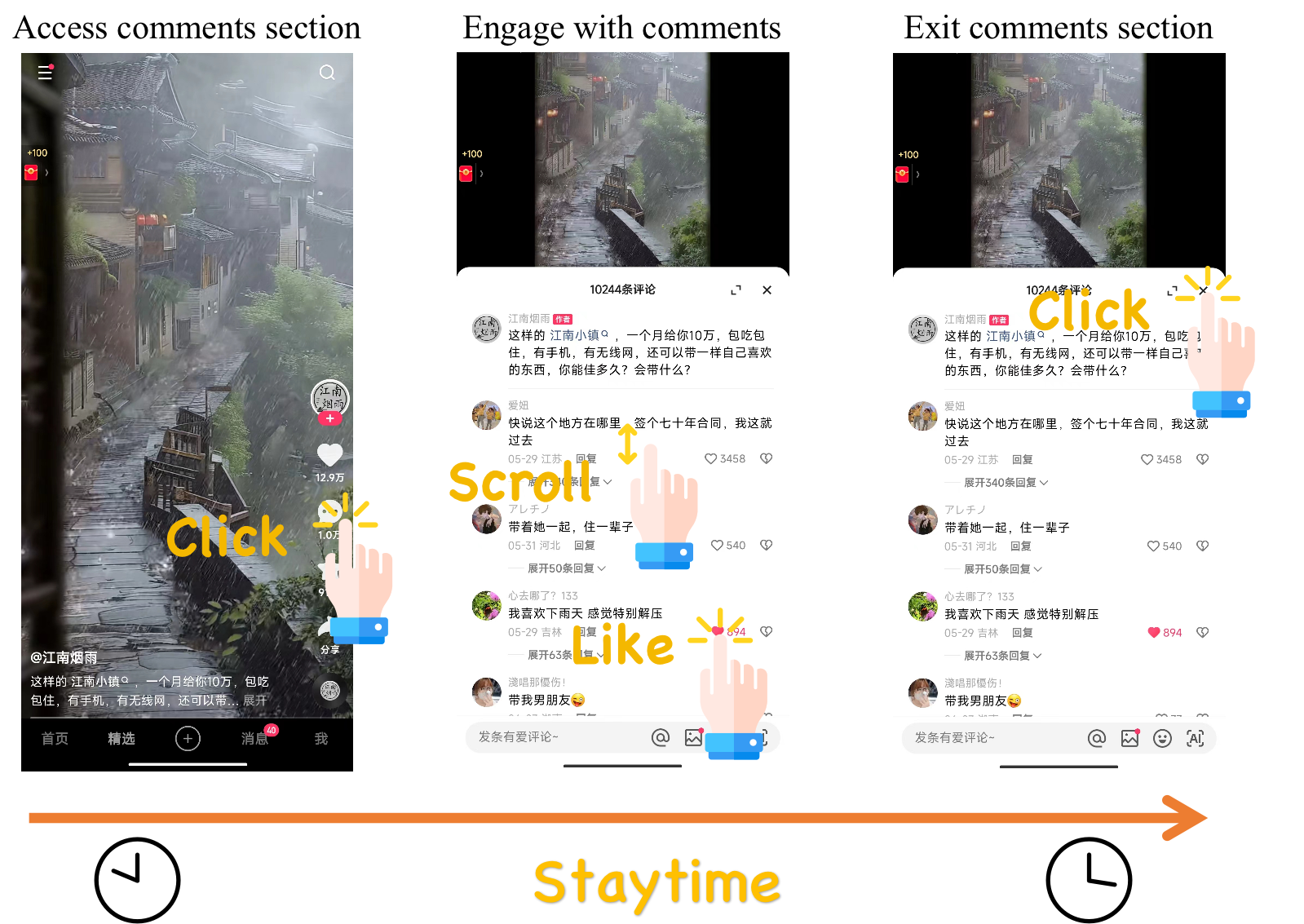}
\vspace{-0.4cm}
\caption{
An illustration of staytime in the comments section. Staytime refers to the total time a user spends in the comments section, starting from the moment they enter until they exit. During this period, users read comments, scroll through, and interact by liking or replying. The figure shows this process, highlighting that staytime encompasses both passive reading and active interaction.}
\vspace{-0.4cm}
\label{illu}
\end{figure}
\begin{figure}[t]
	\centering
 	\subfigure[Relationship between Likes of Top Comments and Staytime] {\includegraphics[width=.9\columnwidth]{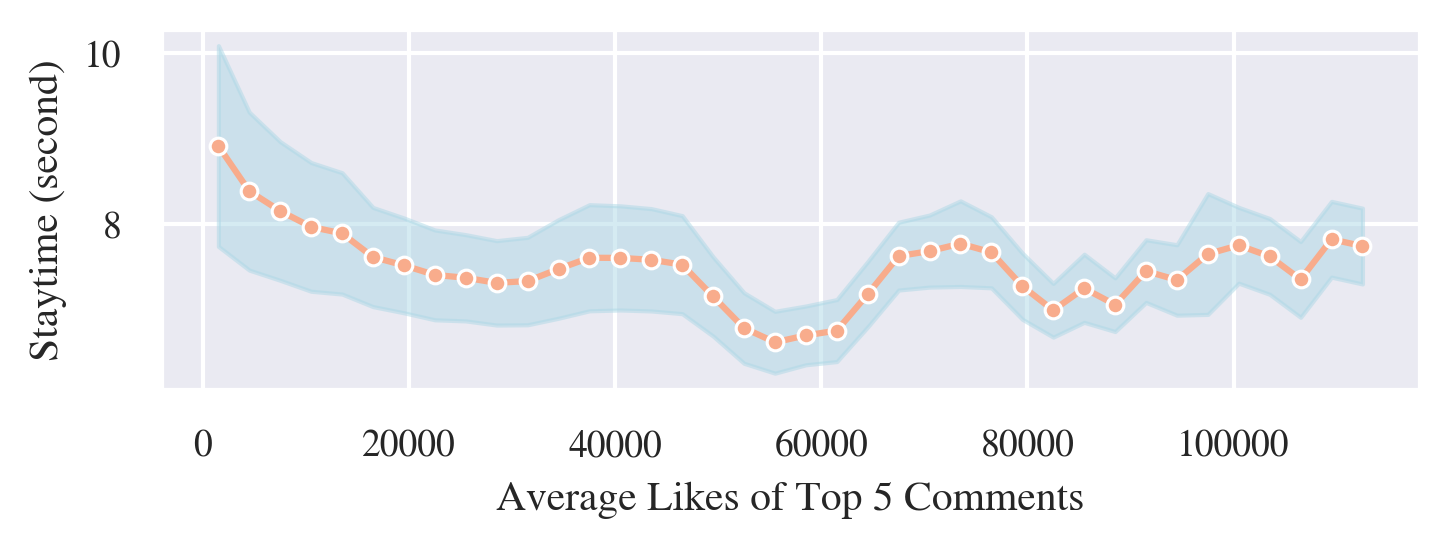}}
	\subfigure[Relationship between Comment Interactions and Staytime] {\includegraphics[width=.9\columnwidth]{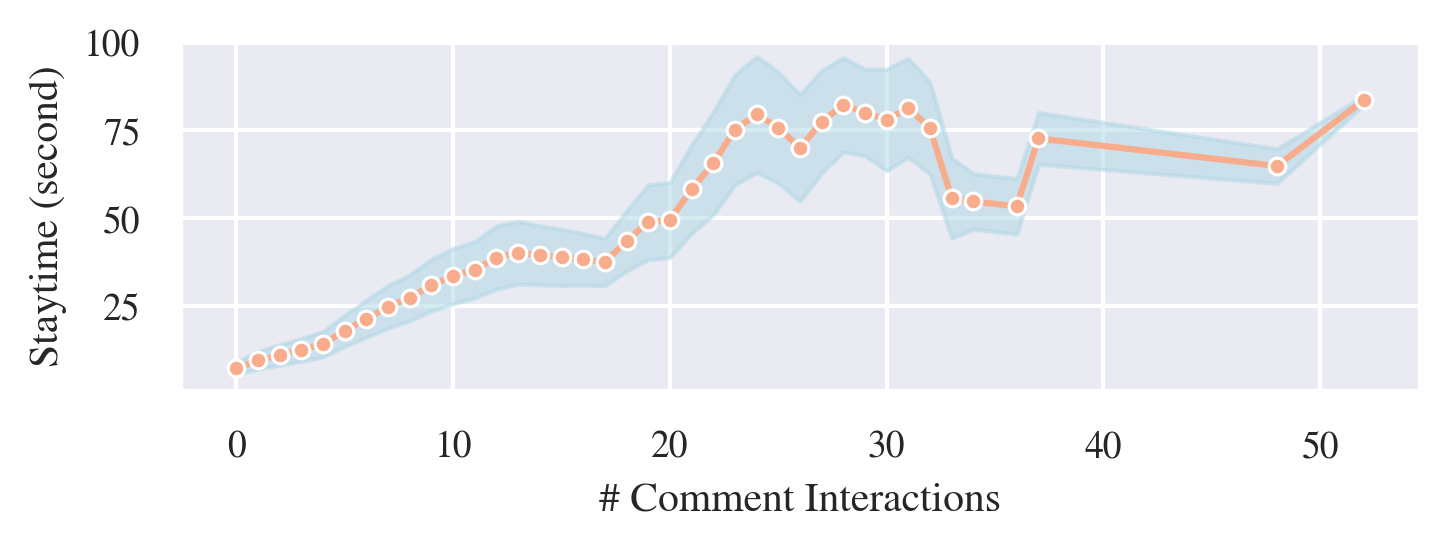}}
	% \subfigure[Relationship between Video Duration and Staytime] {\includegraphics[width=.9\columnwidth]{fig/ana2.png}}
 % 	\subfigure[Relationship between Video Watchtime and Staytime] {\includegraphics[width=.9\columnwidth]{fig/ana3.png}}
\vspace{-0.4cm}
	\caption{Analysis of Staytime on KuaiComt. Data sourced from the KuaiShou App's comments section with a sample size exceeding 10 million. The shaded regions represent the variance within each bucket.}
\vspace{-0.4cm}
	\label{ana}
\end{figure}
\section{Introduction}
% In modern short-video platforms like YouTube, TikTok, and Kuaishou, the comments section has become an essential part of the user experience. Users frequently read and interact with comments, which can influence their overall engagement with the content. One of the key metrics in understanding user behavior within the comments section is \textbf{staytime}—the duration users spend interacting with comments. However, staytime prediction in the comments section remains underexplored, despite its potential to improve the accuracy of recommendation systems and enhance user experience.
In modern short-video platforms like YouTube, TikTok, and Kuaishou, the comments section has become an essential part of the user experience. Users frequently read and interact with comments, which significantly influences their overall engagement with the content. One of the key metrics in understanding user behavior within the comments section is \textbf{staytime}—the total duration users spend from accessing the comments section, reading, and interacting with comments, until they exit. Figure~\ref{illu} illustrates this process, showing the user’s journey through the comments section, from the moment they enter to when they exit. It highlights how staytime encompasses both passive activities, such as reading comments, and active interactions, like scrolling and liking. This comprehensive view of staytime offers valuable insights into user engagement by capturing the full range of behaviors that occur during the user's stay in the comments section. However, staytime prediction in comments sections remains underexplored, despite its potential to improve recommendation systems and enhance user experience.

% 视频停留时长无法对应，但评论区可以对应到...，单点评论停留时长难以获取，但可以获取显示反馈信号
% 多条评论之间相关性，寓意上的关联，整体起作用
% 评论显示信号相对视频播放的隐式信号稀疏

Most existing work focuses on watchtime prediction, which models how long users engage with the video~\cite{covington2016deep, xie2023reweighting, zhan2022deconfounding, zhao2023uncovering, zhao2024counteracting}. However, this approach does not account for the complexities of user interaction within the comments section, which involves multiple comments, varying feedback, and content-related factors. 
% Watchtime prediction methods alone are insufficient to model the nuanced behavior in comments sections.
Unlike video watchtime, which typically cannot be directly attributed to specific content segments, the duration of engagement in the comments section can often be linked to individual comments. This allows for a more granular understanding of user interest and engagement. Although staytime is difficult to attribute to individual comments, interaction signals like likes and replies offer valuable insights into user preferences and behavior. Additionally, the interrelatedness of multiple comments, where the meaning and engagement with one comment might influence the perception of others, plays a significant role in shaping the overall engagement dynamics, highlighting the need for approaches that consider these associative patterns to accurately model user activity in the comments section.

To address this gap, we introduce KuaiComt, a real-world dataset we have constructed and open-sourced. KuaiComt includes user interaction data with both videos and comments, along with rich textual information, such as video titles and comment content. This dataset enables us to study staytime prediction in the comments section. We conducted analytical experiments on KuaiComt and explored three key features, which are illustrated in Fig~\ref{ana}:
\begin{itemize}[leftmargin=*]
% \item The quality of comments, encompassing both their semantic richness and the feedback they receive (e.g., likes, replies), significantly affects user engagement and staytime. Comments that are informative, emotionally charged, or highly relevant to the topic tend to attract more attention, resulting in longer interaction times as users become more immersed in the discussion.

% \item User feedback patterns, such as liking or disliking comments, play a key role in determining staytime. Users typically spend more time on comments they appreciate or find engaging. After thoroughly understanding a comment, they are more likely to interact with it, which leads to increased staytime compared to comments that do not resonate with them.

% \item Interactions with multiple comments do not always have a straightforward, linear impact on staytime. The way users engage with a series of comments can either extend or shorten their overall stay depending on the collective relevance, content quality, and interaction dynamics within the comments section. This complexity highlights the need to consider the overall impact of all comments, rather than focusing solely on individual comment interactions.
\item \textbf{Comment Quality and Staytime:}
   Fig~\ref{ana}(a) shows that staytime decreases as the average likes of the top 5 comments increase, up until around 40,000 likes. This indicates that in less mature comments sections, users spend more time scrolling to find engaging comments. After this point, staytime stabilizes, suggesting users can more easily find relevant content.

\item \textbf{Comment Interactions and Staytime:}
   Fig~\ref{ana}(b) shows that staytime increases steadily with the number of comment interactions, up to around 20 interactions. This demonstrates that users spend more time in the comments section as they engage more with comments, reinforcing the idea that feedback behavior drives higher engagement and longer staytime.

\item \textbf{Non-Linear Relationship Between Multiple Factors and Staytime:}  
   In the second half of both Fig~\ref{ana}(a) and Fig~\ref{ana}(b), the relationship becomes non-linear. In Fig~\ref{ana}(a), after 60,000 likes, staytime fluctuates, showing diminishing returns on engagement. Similarly, in Fig~\ref{ana}(b), after 30 interactions, staytime briefly drops before rising again, indicating that too many interactions may reduce engagement before potentially increasing later. This highlights the complex, non-linear nature of comment interactions and their effect on staytime.
\end{itemize}
These findings highlight the importance of analyzing user behavior in the comments section for accurate staytime prediction. Both comment quality and user interactions significantly influence staytime, with some non-linear characteristics, emphasizing the complexity of user engagement. Understanding these patterns is crucial for improving prediction models and enhancing user experience.

Additionally, given the abundance of textual data available in this scenario, including video titles and comment text, we leverage the powerful semantic understanding of large language models (LLMs)~\cite{achiam2023gpt, touvron2023llama} to enhance our predictions. LLMs can effectively process this rich text information~\cite{liu2021pre, wu2021empowering, zhang2021unbert,yuan2023go, ding2021zeroshotrecommendersystems, zheng2024large}, allowing for deeper insights into comment content and user preferences.

To this end, we propose a two-stage framework \textbf{\ourname~}for staytime prediction. In the first stage, we fine-tune the LLM using a set of domain-specific tasks focused on user behavior in the comments section, including tasks such as staytime bucketing prediction, top comment prediction, and user-comment interaction prediction. This fine-tuning allows the LLM to better understand the context and nuances of user interactions within the comments section.
In the second stage, we integrate the LLM’s embeddings with traditional model features and use two auxiliary tasks—user-agnostic comment ranking (focusing on general comment popularity) and user-specific comment ranking (focusing on individual user preferences)—to predict staytime. These tasks allow the model to capture both general and personalized engagement patterns, offering a more comprehensive approach to staytime prediction.

Our contribution can be summarized as follows:
\begin{itemize}[leftmargin=*]
    \item We are pioneering research on predicting staytime in the comments section, a critical issue in real-world short video recommendation services that has yet to be thoroughly explored.
    \item We introduce a novel two-stage framework for staytime prediction. In the first stage, we fine-tune a LLM using three domain-specific tasks within the comments section. In the second stage, we integrate the LLM with both user-agnostic and user-specific comment ranking tasks to improve staytime prediction.
    \item We have constructed and open-sourced the first real-world video and comment recommendation dataset KuaiComt, which includes user interaction data with both videos and comments, as well as abundant textual information about the videos and comments. Extensive experiments conducted on KuaiComt and online A/B tests have demonstrated the advantages of our framework in staytime prediction tasks across several strong baselines.
\end{itemize}

%% file: sections/3.0_exp.tex
\section{Empirical Study}

\begin{figure*}[t]
\centering
\includegraphics[width=0.85 \textwidth]{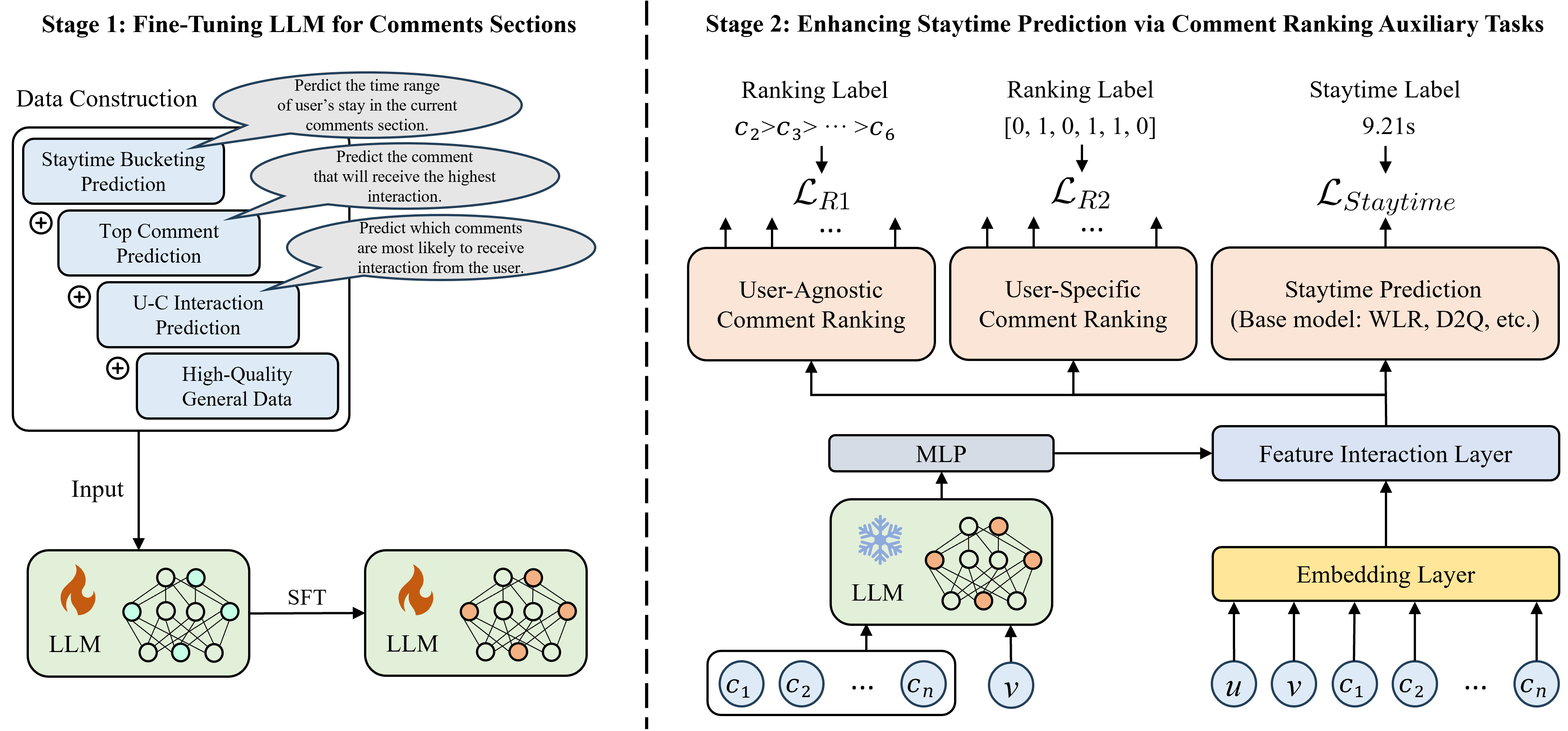}
\vspace{-0.3cm}
\caption{
The overall framework of \ourname. In the first stage, three domain-specific tasks within the comments section are designed for fine-tuning the LLM. In the second stage, embeddings from the LLM for videos and comments are integrated with feature embeddings from traditional models. User-agnostic and user-specific comment ranking auxiliary tasks are utilized to enhance staytime prediction.}
\vspace{-0.35cm}
\label{main_model}
\end{figure*}

In this section, we first present the task definition for predicting the duration of staytime in the comments section. Then, we provide a description and conduct several analyses on our open-sourced real-world dataset, KuaiComt.
\subsection{Task Definition}
In this task, we aim to predict the total staytime \( st_{u,v} \) that a user \( u \) will spend in the comments section of a video \( v \). This prediction is based on the user's interaction history with videos \( S_u = \{v_1, v_2, \dots, v_n\} \), their previous interactions with multiple comments \( S_{c,u} = \{(c_{i,1}, c_{i,2}, \dots, c_{i,k})\} \), and the feature vectors representing the user \( X_u \), the video \( X_v \), and the comments \( X_c = \{X_{c_1}, X_{c_2}, \dots, X_{c_k}\} \) in the video's comments section. The objective is to model the user's engagement by accurately predicting the total staytime in the comments section as:
\begin{equation}
\hat{st}_{u,v} = f(X_u, X_v, X_c).
\end{equation}
By focusing on accurately predicting how long a user will stay in the comments section based on their profiles and the characteristics of the video and multiple comments, we aim to better understand and model user engagement in the comments section of videos.
\subsection{Dataset Description}\label{desp}
Predicting user staytime in the comments section is a relatively new task, and currently, no publicly available datasets exist for this purpose. To fill this gap, we have constructed and open-sourced a large-scale real-world dataset, KuaiComt, collected from the recommendation logs of the video-sharing mobile app, Kuaishou. This dataset includes comprehensive user interaction data with both videos and comments, as well as abundant textual information associated with these videos and comments. KuaiComt is built from the interaction logs of 34,701 users collected between October 1 and October 31, 2023. These logs capture user behaviors such as watching videos, interacting with comments, and engaging in various activities within the platform. The dataset is meticulously designed to provide a robust foundation for developing and evaluating models for video and comment recommendation and prediction tasks. Due to the large number of comment exposures, we have chosen to keep only positive feedback (likes or replies) from users. Additionally, to address privacy and commercial sensitivity concerns, we have implemented data anonymization measures. However, we emphasize that the dataset is constructed based on a comprehensive analysis of user interactions and is aligned with the platform's business strategies. By evaluating our proposed framework on KuaiComt, we aim to demonstrate its effectiveness in predicting user engagement within the comments sections of videos. This dataset offers a valuable resource for researchers and developers seeking to enhance recommendation systems by understanding and modeling user behavior on video platforms. Detailed statistics of KuaiComt are summarized in Appendix~\ref{sta}.
\subsection{Analysis of Staytime on KuaiComt}

% \begin{figure}[t]
% \centering
% \includegraphics[width=0.45 \textwidth]{fig/ana1.png}
% \caption{Relationship between Staytime and Interactions}
% \label{main_model}
% \end{figure}

The staytime within the comments sections of videos significantly influences user engagement and satisfaction. Understanding the factors that drive longer staytime can help in optimizing user experience and enhancing content recommendation strategies.
\subsubsection{Influence of Average Likes of Top Comments}
Fig~\ref{ana}(a) shows the relationship between the average likes of the top 5 comments and the staytime in the comments section. Initially, as the average likes increase, the staytime decreases, suggesting that in less mature comments sections, users need to scroll through more comments to find ones they like. However, as the average likes reach a higher value, the staytime becomes more stable, indicating that users can quickly find engaging comments at the top. This observation highlights how the development of comments sections impacts user behavior and suggests that monitoring the popularity of top comments can help optimize staytime predictions.
\subsubsection{Influence of Users' Interactions}
Fig~\ref{ana}(b) provides a detailed illustration of the relationship between the number of comment interactions (likes or replies) and the staytime within the comments section. Up to around 20 interactions, there is a clear positive correlation, where users tend to spend more time in the comments section as they engage more with comments. Beyond this point, the relationship becomes non-linear, with fluctuations in staytime as interactions increase. This highlights that while more interactions generally increase engagement, there are diminishing returns at higher levels of interaction. Incorporating these fine-grained interaction signals can enhance the accuracy of staytime predictions.
\subsubsection{Influence of Video Watchtime and Video Duration}
Additionally, video duration and watchtime also have significant impacts on staytime in the comments section. The specific patterns and trends observed in these variables suggest that longer videos and extended watchtimes are associated with longer staytime. However, these are not the primary issues addressed in this paper. A detailed analysis of the relationships between these factors and staytime is provided in Appendix~\ref{fa} for further reference in future work.
\subsubsection{Insights}
By analyzing interaction patterns and video characteristics, platforms can more effectively tailor their content and recommendation strategies to enhance user engagement. This approach not only improves user experience but also supports a healthier ecosystem. The focus of this paper is on enhancing staytime predictions through the analysis of fine-grained comment interaction signals. The impact of video duration and watchtime on staytime is reserved for further exploration in future work.

%% file: sections/3_method.tex
\section{Method}
In this section, we describe \ourname~ for the staytime prediction of comments sections in two stages, as illustrated in Figure~\ref{main_model}. The first stage is designed for fine-tuning the LLM through domain-specific tasks in the comments sections, while the second stage utilizes user-agnostic and user-specific comment ranking auxiliary tasks to enhance staytime prediction. 

\subsection{Fine-Tuning LLM for Comments Sections}
we leverage the LLM's exceptional semantic understanding and knowledge reasoning capabilities, particularly in handling comments sections rich with textual information, to fine-tune the LLM. The fine-tuning process involves training the model on three key tasks. These tasks are designed to capture key interaction signals and produce pre-trained embeddings that enhance the model’s ability to make accurate predictions.

\subsubsection{Domain-Specific Data Construction}

In this section, we detail the construction of domain-specific data used to fine-tune the LLM for interactions within comments sections. The data is designed to address three key predictive tasks: \textit{Staytime Bucketing Prediction}, \textit{Top Comment Prediction}, and \textit{User-Comment Interaction Prediction}, each targeting different aspects of user engagement and interaction. These tasks are crucial for training the model to accurately capture and predict user behaviors and preferences in the comments section, thereby enhancing its performance in real-world applications.

\paragraph{Staytime Bucketing Prediction}

This task involves predicting the duration of a user's staytime in the comments section based on their interaction history. Staytime is categorized into different buckets (e.g., brief stay, short stay, moderate stay, long stay). By analyzing patterns in users' past behaviors and current interactions, the LLM learns how these factors influence the length of time a user is likely to spend in the comments section.

\paragraph{Top Comment Prediction}

The objective of this task is to identify which comment in the comments section will attract the highest level of interaction, such as likes or replies. This task helps the LLM understand which types of comments garner the most attention, enabling better comment ranking within recommendation systems.

\paragraph{User-Comment Interaction Prediction}

This task focuses on predicting which comments are most likely to receive interactions (likes or replies) from a specific user based on their current and past behaviors. By modeling user-specific preferences and integrating them with comment-specific features, the LLM generates personalized predictions for interaction likelihoods, contributing to more engaging and interactive comments sections.

\subsubsection{LLM Supervised Fine-Tuning}
To retain the generative capabilities of the large model while enhancing its performance for our comments section's tasks, we fine-tune it using a combination of task-specific and general data~\cite{dong2023abilities}. The fine-tuning dataset comprises data from three domain-specific tasks, alongside additional high-quality general data alpaca-gpt4~\cite{peng2023instruction}. The ratio of these data sources is 1:1:1:3, with the domain-specific tasks contributing equally and the general data being provided in a larger proportion.

The fine-tuning strategy employed is supervised fine-tuning (SFT). This approach allows the model to effectively learn from the constructed domain-specific data while also benefiting from a broader range of general data. The inclusion of general data helps preserve the model's generative capabilities, ensuring it maintains its ability to generate diverse and contextually relevant outputs.

\subsubsection{Pre-trained Embedding Tables Generation}
The final component of stage 1 involves generating pre-trained embedding tables for the video and comments sections. These embeddings encapsulate the patterns learned from the staytime prediction, top comment prediction, and user-comment interaction tasks.
% Specifically, the instruction templates is as follows:
% \noindent \textbf{Video \& Comments Section Prompt:} \textit{Based on the following video title and comments in the comments section , conduct a comprehensive analysis and predict the following: 
% 1) Identify which comment is likely to receive the highest interaction (likes + replies). 
% 2) Analyze which type of user is more likely to interact with each comment. 
% 3) Determine which type of user is more likely to spend a long time in the comments section.  Video Title: [CurVideo]. Comments in the Comments Section: [CommentsList]. This prompt is intended to generate a comprehensive understanding of the video and its comments section to guide subsequent actions.}
% \noindent \textbf{Comment Prompt:} \textit{Based on the following comment text and video title, conduct a comprehensive analysis and predict the following: 
% 1) The richness of the content of the comment and its relevance to the video topic.
% 2) Which type of user is more likely to interact with the comment. 
% Comment Text: [CurComment]. Video Title: [CurVideo]. This prompt is intended to generate an in-depth understanding of the comment to guide subsequent actions.}
For each video and comment, we generate pretrained embeddings and store them in corresponding embedding tables. Specifically, for each video \(v\), we denote the video prompt as \(I^V(v)\), and for each comment \(c\), we denote the comment prompt as \(I^C(c)\). Since next-token prediction is typically the training objective for LLMs, the final token of the entire input sequence captures all the information of that sequence~\cite{neelakantan2022text, wang2023improving}. We extract this embedding as the representation:
\begin{equation}
    e_v = \text{LLM}\left(I^V(v)\right)[-1], \quad e_c = \text{LLM}\left(I^C(c)\right)[-1],
\end{equation}
where \([-1]\) refers to extracting the hidden state of the final token for videos \(I^V(v)\) and comments \(I^C(c)\). The resulting embeddings \(e_v\) and \(e_c\) are stored in the video embedding table \(\mathbf{E}^V\) and the comment embedding table \(\mathbf{E}^C\), respectively. These tables provide enhanced representations for both videos and comments, which are utilized in downstream tasks such as staytime prediction.

\subsection{Enhancing Staytime Prediction via Comment Ranking Auxiliary Tasks}

After generating embedding tables for the videos and comments using the large language model, we will introduce the core components of our \ourname~ framework. In \ourname, besides the standard staytime prediction task, we incorporate two auxiliary tasks related to comment ranking within the comments sections to enhance the training process. These auxiliary tasks are divided into user-agnostic and user-specific comment ranking. The user-agnostic comment ranking task focuses on identifying which comments are likely to become more popular (i.e., receive more likes or replies), while the user-specific comment ranking task predicts which comments the current user is most likely to interact with.

Specifically, after obtaining the embedding tables generated by the large language model for both videos and comments, we first index the video embeddings from the video embedding table $\mathbf{E}^V$ using the video identifiers. Next, we sample comments from the video’s comments section and index the comment embeddings from the comment embedding table $\mathbf{E}^C$ using the comment identifiers. These embeddings are then processed using an MLP (Multi-Layer Perceptron) to ensure dimensional consistency. Subsequently, these embeddings, along with other features processed through the embedding layer, are fed into the feature interaction layer for unified processing. Here, we concatenate them and pass them through a multi-head self-attention layer. This process can be formulated as:
\begin{equation}
e^{\prime}=\text{MHSA}\left(\text{Emb}\left(X_u,X_v,X_c\right)\oplus \text{MLP}\left(\mathbf{E}_{v}^{V}\right)\oplus  \text{MLP}\left(\mathbf{E}_{c_1, c_2, \dots}^C\right)\right),
\end{equation}
where \( u \) denotes the user identifier, \( v \) denotes the video identifier, and \( c_1, c_2, \dots \) denote the comment identifiers. \( X_u \), \( X_v \), and \( X_c \) represent the features of the user, video, and comments within the comments section, respectively. \(\text{MHSA}(\cdot) \) denotes the multi-head self-attention layer, \( \text{Emb}(\cdot) \) denotes the embedding layer, \( \mathbf{E}_v^V \) represents the video embedding obtained by indexing the video embedding table using \( v \), and \( \mathbf{E}_{c_1, c_2, \dots}^C \) represents the comment embeddings obtained by indexing the comment embedding table with \( c_1, c_2, \dots \) sampled from the comments section of \( v \). $\oplus$ denotes the concatenation function.

The resulting representations $e^{\prime}$ can then be fed into standard base staytime prediction models, such as WLR, D2Q, etc., demonstrating the model-agnostic nature of \ourname. The predictions generated by this module are compared with the staytime labels to compute the main loss function, denoted as $\mathcal{L}_{\text{Staytime}}$.

Additionally, since \( e' \) integrates the features of the user, video, and sampled comments, we also use it for the comment ranking tasks. Specifically, to handle both the user-agnostic and user-specific comment ranking tasks, \( e' \) is fed into two separate three-layer MLPs, resulting in two sets of scores, \( \hat{y}^{(1)} \) and \( \hat{y}^{(2)} \), where each set of scores corresponds to the predicted scores for each input comment. This process can be formulated as:
\begin{equation}
\hat{y}^{(1)} = \text{MLP}_{R1}^{(3)}\left(e^{\prime}\right), \quad
\hat{y}^{(2)} = \text{MLP}_{R2}^{(3)}\left(e^{\prime}\right),    
\end{equation}
where \( \text{MLP}_{R1}^{(3)}(\cdot) \) denotes the three-layer MLP for the user-agnostic comment ranking task, and \( \text{MLP}_{R2}^{(3)}(\cdot) \) denotes the three-layer MLP for the user-specific comment ranking task.

Next, we will explain the design and loss calculation process for user-agnostic and user-specific comment ranking task.

\subsubsection{User-Agnostic Comment Ranking Task}
The user-agnostic comment ranking task focuses on identifying which comments are likely to become more popular (i.e., receive more likes or replies). The sampled comments typically have features such as the number of likes, replies, and other engagement metrics. While these features serve as important indicators for the model to learn, directly predicting the exact number of likes or replies can lead to unstable training. To address this, we extend the task into a list-wise ranking problem. Specifically, we use the ListMLE~\cite{xia2008listwise} loss to capture the relative ordering of comments based on their predicted popularity. The loss function \(\mathcal{L}_{R1}\) is defined as:
\begin{equation}
    \mathcal{L}_{R1} = -\log \prod_{i=1}^{n} \frac{\exp\left(\hat{y}^{(1)}_{i}\right)}{\sum_{j=i}^{n} \exp\left(\hat{y}^{(1)}_{i}\right)},
\end{equation}
where \(\hat{y}^{(1)}_{i}\) represents the \(i\) th element of \(\hat{y}^{(1)}\), which corresponds to the predicted score for comment \(c_i\).
\subsubsection{User-Specific Comment Ranking Task}
The user-specific comment ranking task focuses on predicting whether the current user will interact with specific comments within the comment section. This task is crucial for personalizing the content to the user's preferences, as it helps surface comments that are more likely to engage the user. The problem is formulated as a binary classification task, where we estimate the probability that a user will click on or otherwise interact with a particular comment. To model this, we apply a point-wise loss function, specifically the Binary Cross-Entropy (BCE) loss, which is well-suited for such binary prediction tasks. The loss function \(\mathcal{L}_{R2}\) is defined as:
\begin{equation}
    \mathcal{L}_{R2} = -\frac{1}{N} \sum_{i=1}^{N} \left( y_i \cdot \log\left(\hat{y}^{(2)}_{i}\right) + (1 - y_i) \cdot \log\left(1 - \hat{y}^{(2)}_{i}\right) \right),
\end{equation}
where \(\hat{y}^{(2)}_{i}\) represents the \(i\) th element of \(\hat{y}^{(2)}\), which corresponds to the predicted score for comment \(c_i\).
\subsubsection{Formulation of the Total Loss}
After defining the stay-time prediction loss $\mathcal{L}_{\text{Staytime}}$, the user-agnostic comment ranking loss $\mathcal{L}_{R1}$, and the user-specific comment ranking loss $\mathcal{L}_{R2}$, we combine them into a total loss function by performing a weighted sum:
\begin{equation}
    \mathcal{L}_{\text{total}} = \mathcal{L}_{\text{Staytime}} + \lambda_1 \mathcal{L}_{R1} + \lambda_2 \mathcal{L}_{R2},
\end{equation}
where \(\lambda_1\) and \(\lambda_2\) are hyperparameters that control the trade-off between the different loss components. 
% By adjusting these weights, we can fine-tune the model to balance the importance of each task according to the specific needs of the application.

%% file: sections/4_exp.tex
% \newpage
\section{Experiments}
To verify the effectiveness of \ourname, we conduct extensive experiments and report detailed analysis results.
\begin{table}[t]
\centering
\caption{Dataset Statistics.}
  \vspace{-0.4cm}
\begin{tabular}{ccccc}
\hline
\hline \#Users & \#Videos&\#Comments & \#Open-C & \#Inter-C \\
\hline
34,701 & 82,452 & 16,352,904 & 16,033,443 & 1,002,672 \\
\hline
\end{tabular}
\vspace{-0.4cm}
\label{tab:stat}
\end{table}

\subsection{Experimental Setting}
\subsubsection{Dataset}\label{exp:dataset}

\begin{table*}[t]
\centering
\caption{Performance comparison of \ourname~ on different base models in relevance ranking task on KuaiComt. We conducted repeatability experiments and report the average values. The best performance across all models is highlighted in bold. * indicate the improvements over base models are statistically significant ($p$-value < 0.05).}
\vspace{-0.3cm}
\begin{tabular}{l|cccccccc}
\hline
\multirow{2}{*}{Method} & \multicolumn{8}{c}{Metrics}\\
\cline{2-9}
& GAUC & MRR & NDCG@$1$ & NDCG@$3$ & NDCG@$5$ & Staytime@$1$ & Staytime@$3$ & Staytime@$5$\\
\hline
\hline
VR & $0.6611$ & $0.5932$ & $0.4222$ & $0.4146$ & $0.4226$ & $10.1645$ & $9.3845$ & $9.1000$\\
\ourname-VR & $0.6641^{*}$ & $0.6052^{*}$ & $0.4352^{*}$ & $0.4299^{*}$ & $0.4374^{*}$ & $10.5019^{*}$ & $9.7285^{*}$ & $9.4078^{*}$ \\
\hdashline
WLR & $0.6726$ & $0.6242$ & $0.4590$ & $0.4493$ & $0.4542$ & $10.6116$ & $9.8859$ & $9.5337$\\
\ourname-WLR & $\mathbf{0.6746}^{*}$ & $\mathbf{0.6296}^{*}$ & $\mathbf{0.4651}^{*}$ & $\mathbf{0.4558}^{*}$ & $\mathbf{0.4611}^{*}$ & $\mathbf{10.7956}^{*}$ & $\mathbf{10.0433}^{*}$ & $\mathbf{9.6904}^{*}$ \\
\hdashline
NDT & $0.6564$ & $0.6104$ & $0.4429$ & $0.4351$ & $0.4419$ & $10.1040$ & $9.4751$ & $9.2236$ \\
\ourname-NDT & $0.6586^{*}$ & $0.6131^{*}$ & $0.4451$ & $0.4386^{*}$ & $0.4456^{*}$ & $10.1484$ & $9.5477^{*}$ & $9.2969^{*}$ \\
\hdashline
PCR & $0.5839$ & $0.5588$ & $0.3808$ & $0.3833$ & $0.3961$ & $9.1484$ & $8.8004$ & $8.6482$ \\
\ourname-PCR & $0.5864^{*}$ & $0.5623$ & $0.3866$ & $0.3864$ & $0.3975$ & $9.2593$ & $8.8378$ & $8.6381$ \\
\hdashline
D2Q & $0.6675$ & $0.6089$ & $0.4416$ & $0.4318$ & $0.4374$ & $10.2860$ & $9.5572$ & $9.2364$\\
\ourname-D2Q & $0.6707^{*}$ & $0.6189^{*}$ & $0.4522^{*}$ & $0.4442^{*}$ & $0.4502^{*}$ & $10.6242^{*}$ & $9.8692^{*}$ & $9.5281^{*}$\\

\hline
\end{tabular}
\vspace{-0.35cm}
\label{rel}
\end{table*}

\begin{table}[t]
\centering
\caption{Performance comparison of \ourname~ on different base models in staytime prediction task on KuaiComt. 
% We conducted repeatability experiments and report the average values. 
`$\downarrow$' denotes that lower is better for RMSE and MAE, while higher is better for other metrics. 
% The best performance across all models is highlighted in bold. * indicate the improvements over base models are statistically significant ($p$-value < 0.05).
}
\vspace{-0.3cm}
\begin{tabular}{l|cccc}
\hline
\multirow{2}{*}{Method} & \multicolumn{4}{c}{Metrics}\\
\cline{2-5}
& RMSE$\downarrow$ & MAE$\downarrow$ & XGAUC & XAUC\\
\hline
\hline
VR & $8.9727$ & $5.5980$ & $0.5315$ & $0.6058$\\
\ourname-VR & $\mathbf{8.9386}^{*}$ & $5.5511^{*}$ & $0.5357^{*}$ & $0.6076^{*}$\\
\hdashline
WLR & $10.6889$ & $5.8677$ & $0.5340$ & $0.6019$\\
\ourname-WLR & $10.6236^{*}$ & $5.8060^{*}$ & $0.5399^{*}$ & $0.6043^{*}$\\
\hdashline
PCR & $36.1302$ & $14.9512$ & $0.5365$ & $0.5717$ \\
\ourname-PCR & $34.2103^{*}$ & $14.4726$ & $0.5374$ & $0.5732$\\
\hdashline
D2Q & $10.2693$ & $5.0938$ & $0.5467$ & $0.6135$\\
\ourname-D2Q & $10.2500^{*}$ & $\mathbf{5.0721}^{*}$ & $\mathbf{0.5489}^{*}$ & $\mathbf{0.6154}^{*}$ \\
\hline
\end{tabular}
\vspace{-0.35cm}
\label{stp}
\end{table}

% We evaluated the proposed framework on KuaiComt, a large-scale industrial dataset constructed and open-sourced for video and comment recommendation tasks on a short video platform Kuaishou. This dataset includes user interaction data with both videos and comments, as well as abundant textual information about the videos and comments. Specifically, the dataset is built from the interaction logs of 34,701 users collected between October 01 and October 31, 2023. 
We evaluated the proposed framework on the new real-world dataset KuaiComt (described in Section~\ref{desp}). 
To better reflect real-world application scenarios, we filtered out data where the comments section was not opened and only made staytime predictions within the comments section exposure space. We also employed a time-based splitting strategy based on chronological order~\cite{zhao2022revisiting} to divide the dataset. Specifically, to ensure that each user has sufficient historical data for user profiling, we split the data into training, validation, 
and test sets in a 4:1:1 ratio according to the timestamp order. Detailed statistics of the dataset are summarized in Table~\ref{tab:stat}, 
where 
% `\#Impressions-V' refers to the number of times a video is shown to users, 
`\#Open-C' refers to the number of times users open the comments section of a video, and `\#Inter-C' refers to the number of interactions users have with the comments.

\subsubsection{Base Models} 
% In the experiments, we compared the proposed method with the following watchtime prediction baselines for staytime prediction: 
The proposed \ourname~ is model-agnostic, which can be applied to the following watchtime prediction base models for staytime prediction and can improve their performances:
\begin{itemize}[leftmargin=*]
    \item \textbf{VR (Value Regression)} directly fits the observed watchtime using a regression model.
    \item \textbf{WLR~\cite{covington2016deep}} applies weights to samples based on their watchtime.
    \item \textbf{NDT~\cite{xie2023reweighting}} reweights clicks with dwell time by introducing a normalized dwell time function.
    \item \textbf{PCR} converts watchtime into the Play Completion Rate, representing the ratio of the user's watch time to the video's duration.
    \item \textbf{D2Q~\cite{zhan2022deconfounding}} removes duration bias in watch-time prediction by using a causal approach and fitting watchtime quantiles.
\end{itemize}
In our experiments, we applied \ourname~ to these base staytime prediction models, resulting in five versions of our method, referred to as \textbf{\ourname-VR}, \textbf{\ourname-WLR}, \textbf{\ourname-NDT}, \textbf{\ourname-PCR} and \textbf{\ourname-D2Q}.

% To evaluate the generalization capability of our approach and the baseline methods, we incorporate them into various backbone models. In particular, we employ \textbf{FM}~\cite{rendle2012factorization}, \textbf{DCN}~\cite{wang2017deep}, and \textbf{AutoInt}~\cite{song2019autoint} as our backbone models, each of which exemplifies a distinct type of feature interaction: inner product, outer product, and attention mechanisms, respectively.

\subsubsection{Evaluation Metrics}
We evaluated \ourname\ not only for its performance in predicting staytime in comments sections but also for its ranking capabilities, as both are critical in real-world video recommendation scenarios. For staytime prediction, we used the actual staytime \( st_{u,v} \) as the ground truth and employed RMSE (Root Mean Square Error), MAE (Mean Absolute Error), XGAUC, and XAUC~\cite{zhan2022deconfounding} as evaluation metrics. For relevance ranking based on user interest, following D$^2$Co~\cite{zhao2023uncovering}, we defined a positive sample as a user staying in the comments section for an extended time, and a negative sample otherwise. Specifically, 
\begin{equation}
r_{u,v} = 
\begin{cases}
1 & \text{if }st_{u,v} > st_{0.7}, \\
0 & \text{otherwise,}
\end{cases}
\end{equation}
where \( st_{0.7} \) represents the $70\%$ percentile of the observed staytime, which is considered the threshold for determining a long stay in the comments section. \( r_{u,v} \) is used as the ground truth for evaluating the relevance ranking task, with GAUC, MRR, NDCG@$1$, NDCG@$3$, NDCG@$5$, Staytime@$1$, Staytime@$3$ and Staytime@$5$ serving as evaluation metrics, with Staytime@$n$ representing the average staytime of the top $n$ ranked videos after sorting.

\subsubsection{Implementation Details}
For the implementation of the baselines, both WLR and NDT utilize a dual-tower model. In WLR, one tower is dedicated to determining whether a user opens the comment section, making it applicable for relevance ranking tasks. However, since the model's prediction cannot be reversed into staytime using an inverse transformation function in NDT, it is only used for relevance ranking. PCR and D2Q are designed using transformation functions for both relevance ranking and watch time prediction. For relevance ranking, candidate videos are ranked based on the prediction scores generated by the models trained with these methods. For staytime prediction, we first convert the model's predictions into staytime using their inverse transformation functions. For the implementation of D2Q, we divided the videos into 30 equal buckets based on their duration. In the implementation of our framework \ourname, for each data point, the number of sampled comments is 6. The sampling range includes the top 7 popular comments and all comments interacted with by users in the current comments section. For the large language model, we selected Qwen2-7b\footnote{https://github.com/QwenLM/Qwen2}~\cite{yang2024qwen2} and trained it for one epoch on a dataset of 15,000 samples, using low-rank adaptation based on LoRA~\cite{hu2021lora}. Hyperparameters \(\lambda_1\) and \(\lambda_2\) are selected from \{$1e^{-4}$, $1e^{-3}$, $1e^{-2}$, $1e^{-1}$, $1$\} and we carefully search hyperparameters for optimal performance.

\subsection{Overall Performance}
\subsubsection{Relevance Ranking Task}
For the task of relevance ranking evaluation based on user interest prediction, Table~\ref{rel} highlights several important observations. WLR stands out as the best-performing baseline across all metrics, including GAUC, MRR, NDCG, and Staytime. This indicates that its method of applying weights to watch time effectively captures user engagement in the comments section, making it particularly suitable for staytime prediction tasks. 
% Additionally, its dual-tower architecture allows for direct optimization through ranking metrics, enabling it to surpass other models. 
D2Q demonstrates the second-highest performance among the baseline models, largely due to its approach of bucketizing videos by duration and estimating staytime based on these groupings. This suggests that accounting for the influence of video duration on staytime contributes significantly to performance improvements. In contrast, PCR consistently demonstrates the weakest performance across all metrics. While PCR is effective for video watchtime prediction, it performs poorly in predicting staytime. This is because PCR primarily captures the ratio of watchtime to duration, and using this ratio to estimate staytime lacks clear practical relevance in the context of user engagement with the comments section.

Across all base models, our model-agnostic framework, \ourname, demonstrates consistent performance improvements. When applied to these base models, \ourname\ shows significant gains in all metrics. This suggests that \ourname’s ability to leverage LLMs and fine-grained comment interaction signals helps enhance not only the ranking of relevant content but also the accurate prediction of how long users are likely to engage with comments. The fact that \ourname\ improves both weak models like PCR and strong models like WLR shows its versatility and robustness across different prediction strategies. This underlines \ourname’s potential to be applied across diverse scenarios and models, making it highly adaptable and effective in predicting user behavior in the comments section. 
\subsubsection{Staytime Prediction Task}
Table~\ref{stp} shows several key insights into the performance of different models in staytime prediction. D2Q stands out as the best-performing model, showing the optimal MAE, XGAUC, and XAUC, despite a slightly higher RMSE compared to VR. As mentioned earlier, its method of grouping videos based on duration has proven effective in improving staytime prediction. In contrast to the ranking task, the WLR model performs slightly less effectively, possibly due to errors introduced by its dual-tower architecture in the prediction process. However, PCR remains the weakest across all metrics, indicating its struggles with accurate staytime predictions, as previously mentioned.

Across all base models, \ourname\ shows improvements by reducing error metrics (RMSE and MAE) and increasing XGAUC and XAUC scores. These improvements demonstrate \ourname’s adaptability in enhancing staytime prediction performance.
\begin{table}[t]
\centering
\caption{Ablation study of \ourname~on KuaiComt.}
% \caption{Ablation study on four metrics of \ourname-WLR and \ourname-D2Q for both relevance ranking task and staytime prediction task on KuaiComt.}
\vspace{-0.3cm}
\begin{tabular}{l|cccc}
\hline
\multirow{2}{*}{Method} & \multicolumn{4}{c}{Metrics}\\
\cline{2-5}
& NDCG@$5$ & Staytime@$5$ & MAE$\downarrow$ & XAUC\\
\hline
\hline\ourname-WLR & $0.4611$ & $9.6904$ & $5.8060$ & $0.6043$\\
w/o $SFT$ & $0.4562$ & $9.5427$ & $5.8536$ & $0.6036$\\
w/o $\mathcal{L}_{R1}$ & $0.4590$ & $9.6638$ & $5.8904$ & $0.6036$\\
w/o $\mathcal{L}_{R2}$ & $0.4595$ & $9.6726$ & $5.9519$ & $0.6040$\\
\hdashline
\ourname-D2Q & $0.4502$ & $9.5281$ & $5.0821$ & $0.6154$ \\
w/o $SFT$ & $0.4467$ & $9.4995$ & $5.0933$ & $0.6142$\\
w/o $\mathcal{L}_{R1}$ & $0.4485$ & $9.5090$ & $5.0920$ & $0.6151$\\
w/o $\mathcal{L}_{R2}$ & $0.4491$ & $9.5125$ & $5.1372$ & $0.6138$\\
\hline
\end{tabular}
\vspace{-0.3cm}
\label{abl}
\end{table}
\subsection{Ablation Study}
To explore how the proposed techniques affect overall performance, we conducted an ablation study on relevance ranking and staytime prediction tasks. Specifically, we examined four variants of the two best-performing models, \ourname-WLR and \ourname-D2Q: (1) w/o $SFT$, which removes the supervised fine-tuning stage of the large model, directly outputting the embedding representation. (2) w/o $\mathcal{L}_{R1}$, which removes the user-agnostic comment ranking auxiliary task. (3) w/o $\mathcal{L}_{R2}$, which removes the user-specific comment ranking auxiliary task. The results shown in Table ~\ref{abl} clearly indicate that removing any of these components leads to a decline in performance. For \ourname-WLR, the absence of $SFT$ and $\mathcal{L}_{R1}$ has a more pronounced negative impact, particularly on Staytime@5 and MAE. Removing $SFT$ reduces the model’s ability to optimize embeddings based on task-specific data, resulting in poorer predictions for both staytime and ranking accuracy. Similarly, removing $\mathcal{L}_{R1}$ weakens the model's ability to personalize rankings based on user-comment preferences, leading to performance drops. In \ourname-D2Q, removing $SFT$ also leads to a significant decline in all metrics, as the model loses its ability to adjust embeddings during fine-tuning, which is crucial for capturing subtle relationships in the data. While removing $\mathcal{L}_{R1}$ and $\mathcal{L}_{R2}$ has a similar impact, the overall trend confirms that each proposed technique plays an important role in enhancing the model’s effectiveness across different tasks.
% \subsubsection{Effectiveness of Supervised Fine-Tuning}
% \subsubsection{Effectiveness of User-Agnostic Comment Ranking}
% \subsubsection{Effectiveness of User-Specific Comment Ranking}

\subsection{Further Analysis}
\begin{figure}[t]
\centering
\includegraphics[width=0.45 \textwidth]{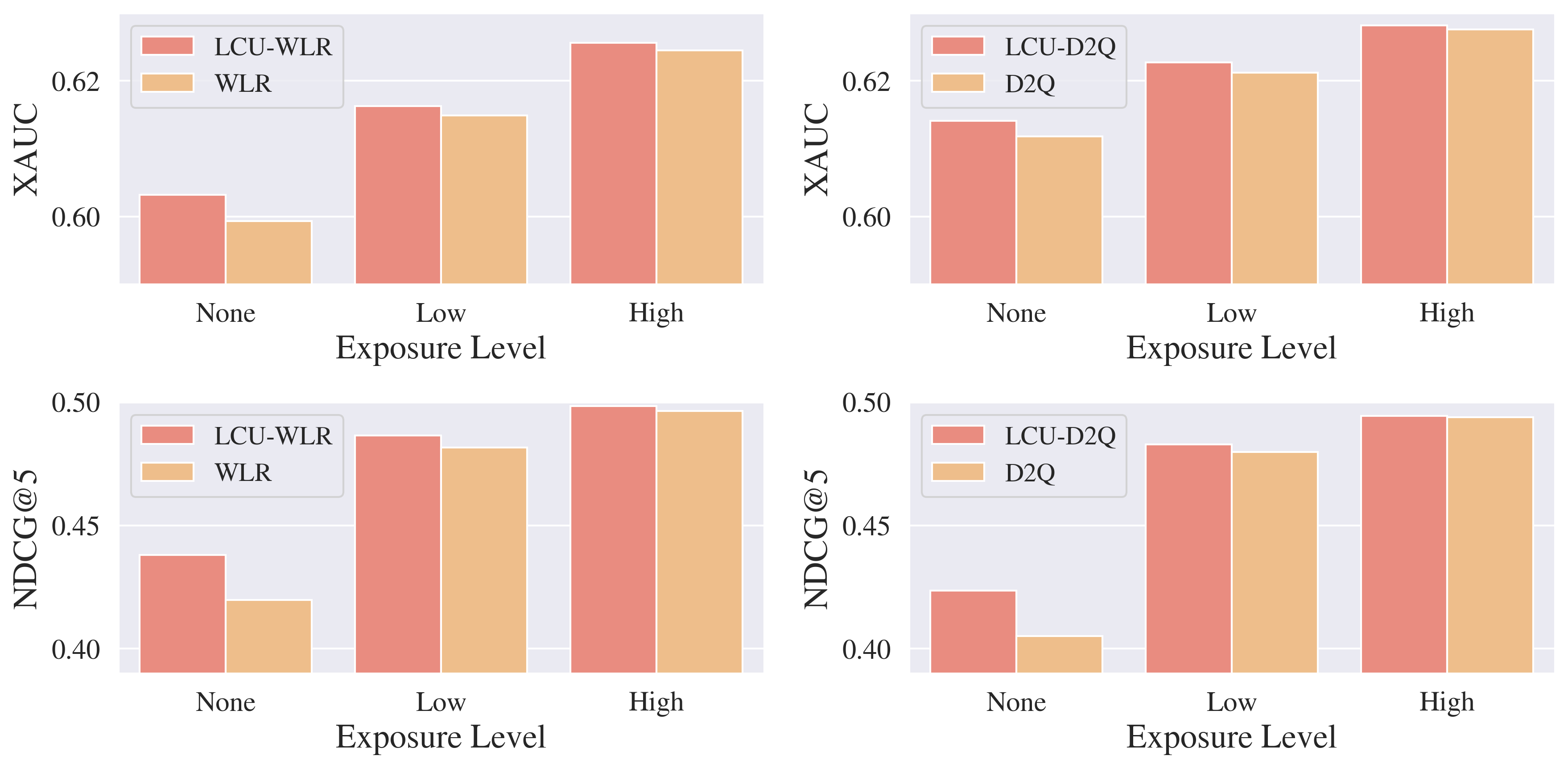}
\vspace{-0.4cm}
\caption{Analysis of \ourname~across different exposure level groups.}
\vspace{-0.4cm}
\label{ae}
\end{figure}
\begin{figure}[t]
\centering
\includegraphics[width=0.45 \textwidth]{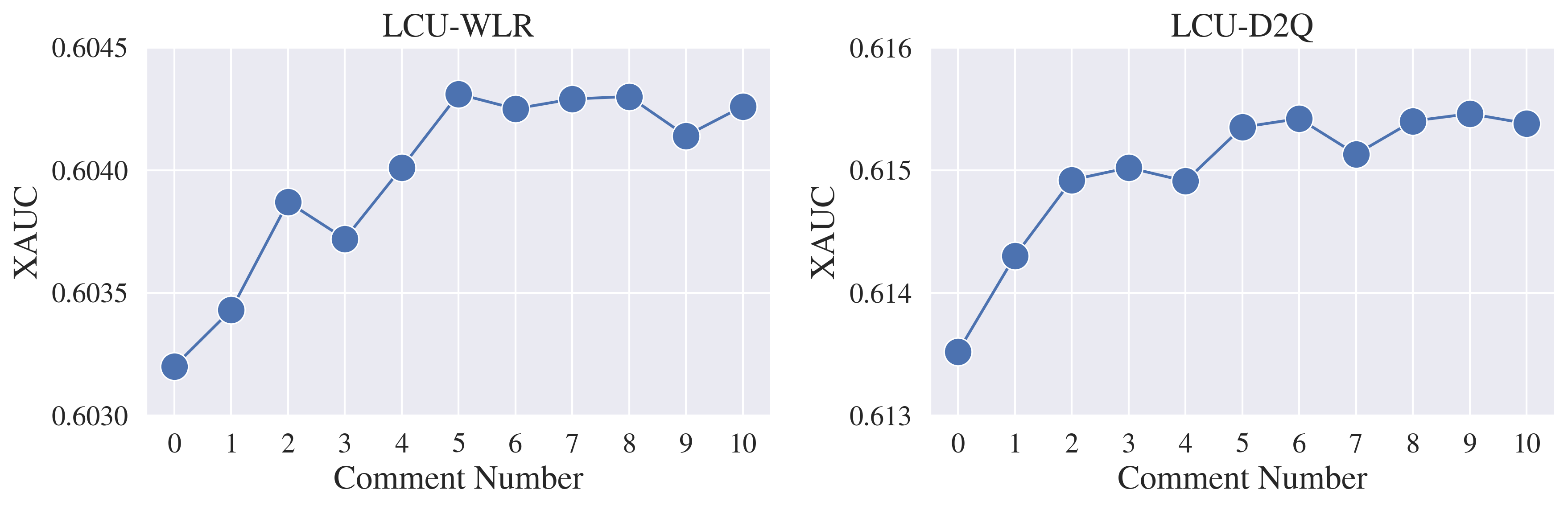}
\vspace{-0.4cm}
\caption{Analysis of comment number impact on \ourname.}
\vspace{-0.4cm}
\label{ae2}
\end{figure}
\subsubsection{Effectiveness of LLM's embedding in Enhancing Cold-Start Video Performance} We evaluated the effectiveness of \ourname~ in improving the ranking and prediction performance of cold-start videos by comparing it with base models WLR and D2Q. The training dataset was divided by exposure frequency, and the test set was categorized into three groups: "None", "Low" and "High", representing videos with no, low, and high exposure in the training set, respectively. As shown in Figure~\ref{ae}, \ourname~ significantly outperforms the baseline models, especially for cold-start videos (None group), in both XAUC and NDCG@5 metrics.
The superior performance of \ourname~ highlights the value of incorporating LLM-generated embeddings, which capture relevant content and contextual information. These embeddings help the model make more accurate predictions, even for videos with little or no interaction history, leading to better ranking precision and an improved user experience.
\subsubsection{Impact of Comment number on \ourname}
We analyzed the effect of different comment numbers on XAUC performance in the comment ranking auxiliary tasks. The results in Figure~\ref{ae2} show that as the number of comments increases, the XAUC improves steadily, with a notable rise from $0$ to $6$ comments before stabilizing. This indicates that more user interaction through comments provides the model with richer information, enhancing its ability to make more accurate predictions and improving overall ranking performance.
% \subsubsection{Impact of $\lambda_2$}
\subsection{Online A/B Testing}
\begin{figure}[t]
\centering
\includegraphics[width=0.45 \textwidth]{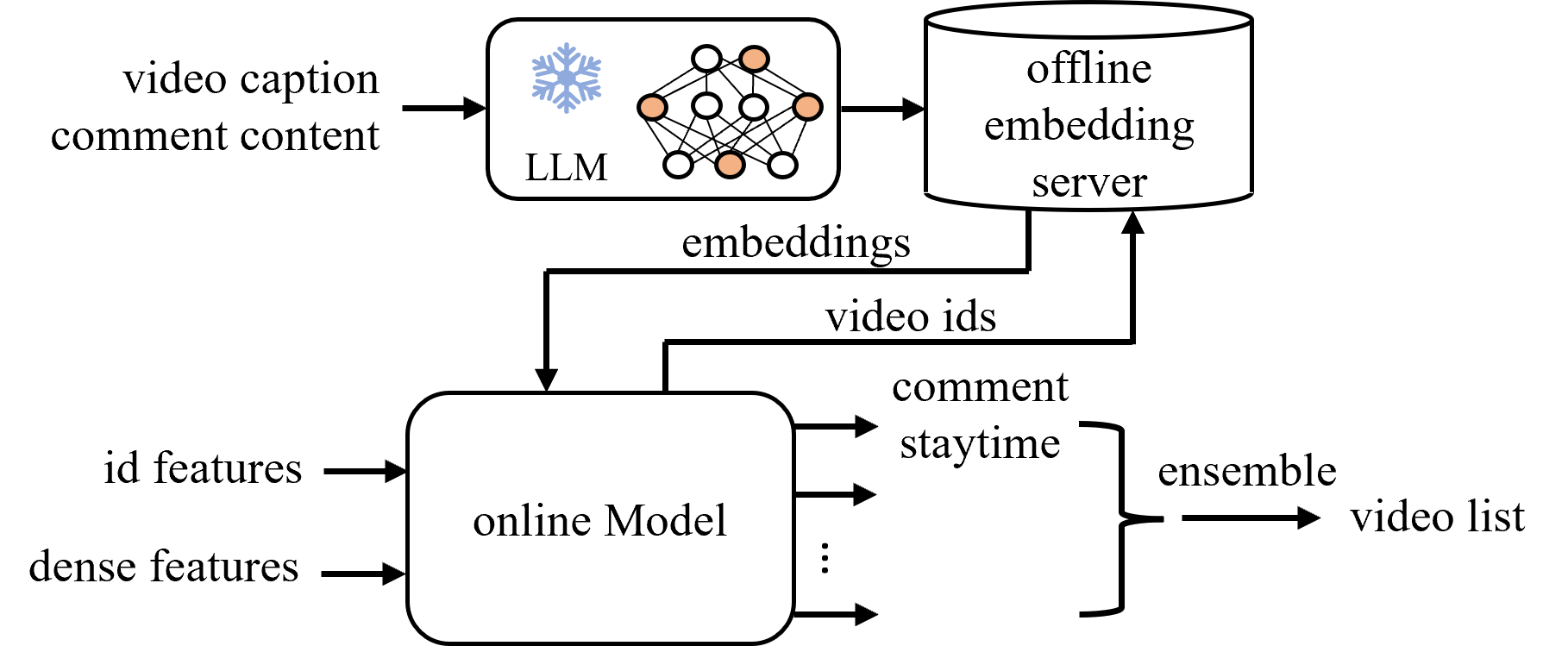}
\vspace{-0.3cm}
\caption{Workflow for the online deployment of \ourname.}
\vspace{-0.3cm}
\label{online_workflow}
\end{figure}

To further validate the effectiveness of \ourname, we conducted a two-week online A/B test on the Kuaishou platform. We integrated our method into the existing recommendation workflow for comparison, as illustrated in Figure~\ref{online_workflow}. Due to the high cost of LLMs and the large number of candidate videos in real-world applications, we sampled 150,000 high-popularity videos and fine-tuned the LLM offline using their comments. The LLM-generated embeddings were then stored in an embedding server for online usage. These LLM-enhanced embeddings were incorporated into an online multi-objective model, with comment staytime being one of the factors influencing the final recommendation. We used two key metrics to measure user engagement: (1) Staytime: the average staytime spent in the video comments section per user. (2) Exposure Num.: the average number of comments exposed per user. The results, shown in Table~\ref{online}, reveal that \ourname~ achieved significant improvements in both staytime and exposure number, highlighting its strong potential for real-world deployment on video platforms.
\begin{table}[t]
\centering
\caption{Results of online A/B testing on KuaiShou.}
\vspace{-0.3cm}
\begin{tabular}{lr}
\hline
\textbf{Online Metrics} & \textbf{Relative improvement} \\ \hline
Staytime              & +1.27\%      \\ 
Exposure Num.        & +0.81\%      \\ \hline
\end{tabular}
\vspace{-0.3cm}
\label{online}
\end{table}

%% file: sections/2_related.tex
\section{Related Work}
\subsection{Watchtime Prediction}
% In recent years, predicting watch time has become a crucial focus in video recommendation systems, serving as a key metric for assessing user engagement. Early efforts, such as the Weighted Logistic Regression (WLR) model~\cite{covington2016deep} used in YouTube's recommendation system, aimed to predict watch time but encountered significant challenges, particularly with biases linked to video duration. To overcome these limitations, newer methods like D2Q~\cite{zhan2022deconfounding}, D2Co~\cite{zhao2023uncovering} and CWM~\cite{zhao2024counteracting} have been developed to correct for duration bias and handle noisy viewing data, resulting in more accurate predictions. Additionally, multi-task learning techniques~\cite{wang2024home} and advanced modeling approaches~\cite{tang2020progressive} have been introduced to further enhance the precision of watch time predictions by addressing the complexities of user behavior and interactions with content. 
% % These innovations reflect a growing emphasis on refining watch time prediction to improve the overall effectiveness of video recommendation systems.
% However, directly applying watchtime prediction to comment staytime is inadequate, as it overlooks the complexity of user engagement in the comments section, where interactions like likes, replies, and the interrelation between multiple comments provide deeper insights into user behavior and preferences.
In recent years, watchtime prediction has become a key focus in video recommendation systems to measure user engagement. Early models like WLR~\cite{covington2016deep} aimed to predict watch time but struggled with biases related to video length. Newer methods, such as D2Q~\cite{zhan2022deconfounding}, D2Co~\cite{zhao2023uncovering}, and CWM~\cite{zhao2024counteracting}, were developed to address these biases and handle noisy data, improving prediction accuracy. Advanced techniques like multi-task learning~\cite{wang2024home} and other modeling approaches~\cite{tang2020progressive} further enhance prediction by capturing complex user behaviors.
However, applying watchtime prediction to comment staytime is inadequate, as it ignores the complexity of user interactions in the comments, such as likes, replies, and comment relationships, which provide deeper insights into user preferences.

\subsection{LLMs for Recommendation}
Inspired by the advancements of large language models (LLMs) like GPT4~\cite{achiam2023gpt} and LLaMA~\cite{touvron2023llama}, recent studies have explored their application in recommendation systems. LLMs are used either as text encoders to generate embeddings for traditional models~\cite{liu2021pre, zhang2021unbert, wu2021empowering, harte2023leveraging, hou2022towards, ren2024representation} or as standalone models that leverage their pre-trained knowledge~\cite{bao2023tallrec, hegselmann2023tabllm, kang2023llms, hou2024bridging, xu2024prompting, zhang2023recommendation} for tasks such as zero-shot and few-shot recommendations~\cite{brown2020languagemodelsfewshotlearners}. For instance, MoRec~\cite{yuan2023go} and ZESRec~\cite{ding2021zeroshotrecommendersystems} utilize LLMs to create alternative item representations, while Recformer~\cite{li2023text} integrates them for holistic text encoding. 
% However, these methods often face challenges like increased computational complexity and alignment issues between embeddings and textual meanings. Despite these challenges, 
These innovations show promise in adapting LLMs to various recommendation tasks with minimal fine-tuning.

%% file: sections/5_conc.tex
\section{Conclusion}
% In this paper, we introduced the staytime prediction problem in the context of short-video platform comment sections, emphasizing its importance for understanding user engagement and enhancing user experience. To address this challenge, we open-sourced KuaiComt, the first real-world dataset specifically designed for studying staytime in comment sections. Additionally, we proposed \ourname, a model-agnostic framework that leverages large language models (LLMs) to process rich textual information and incorporate fine-grained comment ranking signals. By fine-tuning LLMs for domain-specific comment behavior and integrating them with traditional models, \ourname~ demonstrates significant improvements in staytime prediction, as validated by both extensive offline experiments and real-world A/B testing. Our work establishes a strong foundation for further exploration of staytime prediction, with potential applications in optimizing recommendation systems and enhancing user engagement across various platforms.
In this paper, we introduced the staytime prediction problem for short-video platform comment sections, emphasizing its role in understanding user engagement. We released KuaiComt, the first dataset for studying comment staytime, and proposed \ourname, a framework that combines large language models (LLMs) with traditional models for improved staytime prediction. By fine-tuning LLMs for comment understanding and comment ranking tasks, \ourname~ demonstrated significant improvements, validated by offline experiments and real-world A/B testing. Our work provides a foundation for future research in staytime prediction and its application in optimizing recommendation systems and enhancing user engagement.

%% file: sections/6_app.tex
% \section{KuaiComt}
% \subsection{Related Work}
% \subsubsection{Existing Research Topics}
\begin{table}[t]
\centering
\caption{Dataset Statistics of KuaiComt.}
  \vspace{-0.3cm}
\begin{tabular}{ccc}
\hline
\hline \#Users & \#Videos&\#Comments\\
\hline
34,701 & 82,452 & 16,352,904\\
\hline
\hline
\#Impressions-V & \#OpenComments-V & \#Interactions-C\\
\hline
119,696,682 & 16,033,443 & 1,002,672 \\
 \hline
\end{tabular}
\label{staref}
\end{table}

\begin{table*}[t]
\centering
\caption{Brief Descriptions of KuaiComt Features.}
  \vspace{-0.3cm}
\begin{tabular}{l|l}
\hline
\hline Feature & Brief Descriptions\\
\hline
User feature & Users have abundant side information, e.g., user active degree, follow count.\\
\hline
Video feature & Videos have abundant side information, e.g., caption, duration.\\
 \hline
 Comment feature & Comments have abundant side information, e.g., comment content, comment like cnt.\\
 \hline
 V-inter feature & Video-interactions have 12 features, e.g., comment stay time, play time, likes, and follows.\\
 \hline
 C-inter feature & Comment-interactions has 2 features, including 2 types of user feedback: likes and replies.\\
 \hline
\end{tabular}
\label{ft}
\end{table*}

\begin{figure}[t]
	% \centering

	\subfigure[Relationship between Video Duration and Staytime] {\includegraphics[width=.9\columnwidth]{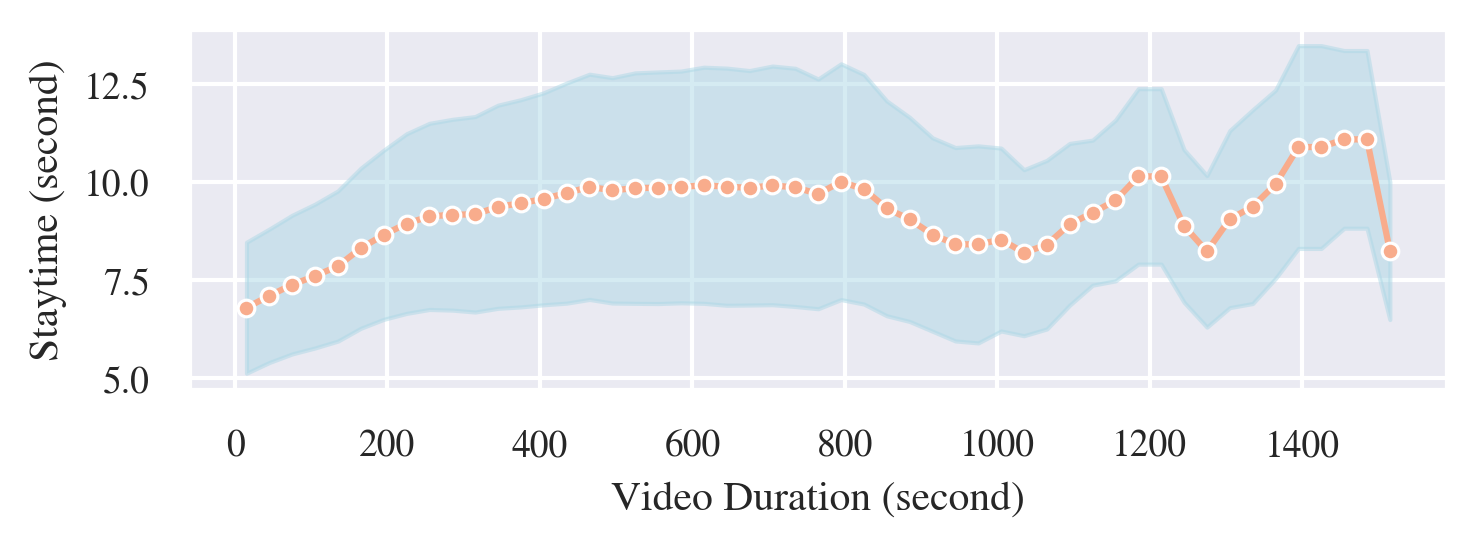}}
 
 	\subfigure[Relationship between Video Watchtime and Staytime] {\includegraphics[width=.9\columnwidth]{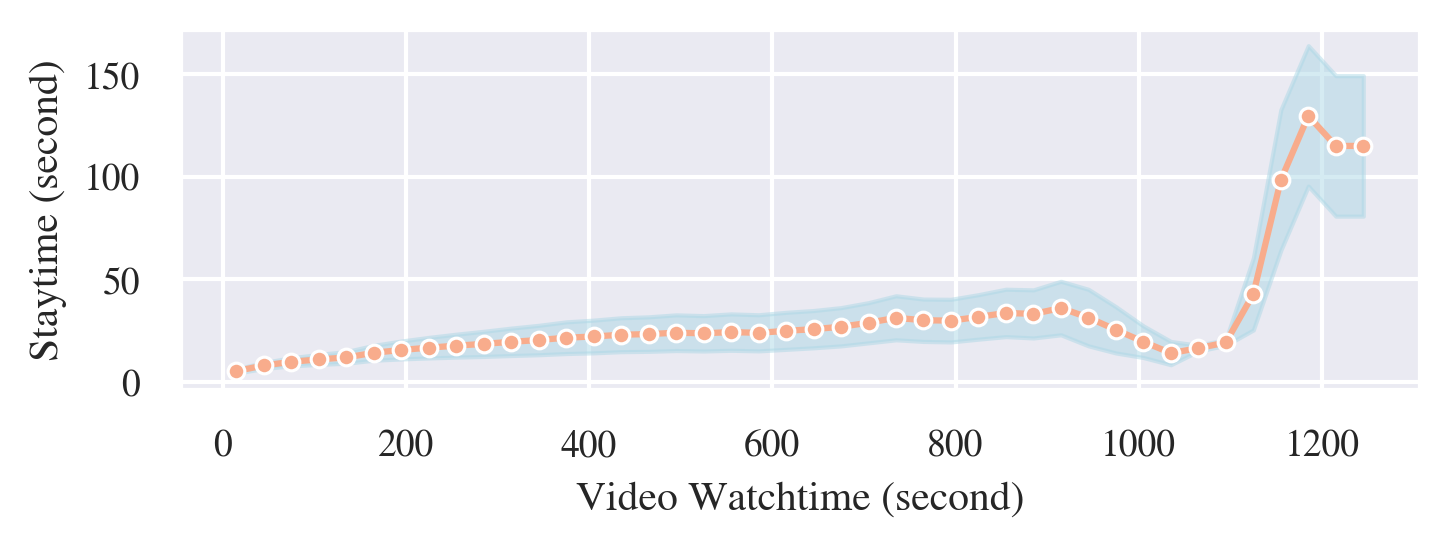}}
  
	\caption{Analysis of Staytime on KuaiComt. Data sourced from the KuaiShou App's comments section with a sample size exceeding 10 million. The shaded regions represent the variance within each bucket.}
	\label{anaref}
\end{figure}

\section{Detailed Statics of KuaiComt}\label{sta}
KuaiComt contains the real behavior of 34,701 users on the Kuaishou app from September 30, 2023, to November 3, 2023. Due to the large number of comment impressions to users, we only provide data on user interactions with comments (likes and replies). Videos with fewer than 55 comments and comments with fewer than 2 interactions were filtered out. Additionally, video titles and comment texts were anonymized. The detailed statics are summarized in Table~\ref{staref}, where `Impressions-V' denotes the impressions of videos to users, `OpenComments-V' denotes the behavior of users opening the comments section, and `Interactions-C' denotes user interactions with comments (such as likes or replies). The short descriptions for each feature filed are listed in Table~\ref{ft}. Please visit our website for more details and examples.
\section{Further Analysis on KuaiComt}\label{fa}
Fig~\ref{anaref}(a) shows the relationship between video duration and staytime. We observe that as video duration increases up to around 600 seconds, the staytime gradually increases, suggesting that longer videos encourage more engagement in the comments section. However, after 600 seconds, the staytime fluctuates, indicating that for very long videos, the impact on staytime becomes less predictable.

In Fig~\ref{anaref}(b), the relationship between video watchtime and staytime is presented. The staytime steadily increases as watchtime approaches 1200 seconds, with a sharp increase observed beyond this point. This suggests that users who watch longer portions of a video tend to spend more time in the comments section, with a notable spike in engagement when users have watched most or all of the video. However, after 1200 seconds, the staytime shows slight fluctuations, which may indicate variations in user engagement based on video content or other factors.